\documentclass[a4paper,10pt]{article}
\usepackage{amsfonts}
\usepackage{amsmath}
\usepackage{amssymb}
\hoffset= - 1.0in         % switch off for draft style
\begin{document}

\begin{titlepage}
\hfill ITEP/TH-09/09

\vspace{5cm} \centerline{\begin{Huge}Degenerate Sklyanin
Algebras\end{Huge}}

\vspace{1cm} \centerline{\begin{Large}Andrey Smirnov
\end{Large}} \vspace{0.5cm}
\centerline{\begin{Large}\texttt{E-mail:
asmirnov@itep.ru}\end{Large}} \vspace{5 mm}
\centerline{\begin{Large} Moscow Institute of Physics and
Technology, Moscow, Russia \end{Large}} \vspace{5 mm}
 \centerline{\begin{Large}
Institute for Theoretical and Experimental Physics,  Moscow,
Russia
\end{Large}}
\vspace{1cm} \centerline{\textbf{Abstract}} New trigonometric and
rational solutions of the quantum Yang-Baxter equation (QYBE) are
obtained by applying some singular gauge transformations to the
known Belavin-Drinfeld elliptic R-matrix for $sl(2,\mathbb{C})$.
These solutions are shown to be related to the standard ones by
the quasi-Hopf twist. We demonstrate that the quantum algebras
arising from these new R-matrices can be obtained as special
limits of the Sklyanin algebra. A representation for these
algebras by the difference operators is found. The
$sl(N,\mathbb{C})$-case is discussed.

\end{titlepage}
\section{Introduction}
In \cite{LOZ} the equivalence between the elliptic N-particle
Calogero-Moser (CM) system and the elliptic Euler-Arnold top was
demonstrated on the base of Hitchin approach to integrable
systems. Within this framework the Lax operators arise naturally
as the Higgs fields associated with holomorphic bundles of
different degrees over an elliptic curve $E_{\tau}$. The relation
between the Lax operators has the form of a gauge transformation:
\begin{equation}
\label{SHC}
 L^{top}_{ell}(z,\tau)=\Xi(z)\,L^{CM}_{ell}(z,\tau)\,\Xi^{-1}(z)
\end{equation}
where the elements of the matrix $\Xi(z)$ are expressed through
theta-functions depending on the coordinates of CM particles. The
trigonometric and rational degenerations of (\ref{SHC}) is not
straightforward and careful analysis is needed. For example, the
naive trigonometric limit of (\ref{SHC}) leads to divergent
expressions because the matrix elements of $\Xi(z)$ are singular
in this limit. In \cite{Smir1,Smir2} we overcame this difficulty
by applying a singular gauge transformation depending on $\tau$ to
(\ref{SHC}) before taking the limit:
$$
A(\tau)\,L^{top}_{ell}(z,\tau)\,A^{-1}(\tau)=A(\tau)\,\Xi(z)\,L^{CM}_{ell}(z)\,\Xi^{-1}(z)\,A^{-1}(\tau).
$$
As a result we obtained the Lax operator for the trigonometric top
$L^{top '}_{trig}(z)$ which differs from the standard
trigonometric degeneration of  the elliptic Lax operator $L^{top
}_{trig}(z)$:
$$
L^{top
'}_{trig}(z)=\lim\limits_{\tau\rightarrow\,\textrm{i}\infty}\,
A(\tau)\,L^{top}_{ell}(z,\tau)\,A^{-1}(\tau) \ \ \ \ L^{top
}_{trig}(z)=\lim\limits_{\tau\rightarrow\,\textrm{i}\infty}\,
\,L^{top}_{ell}(z,\tau)
$$
It is well known that the quantization of phase space of the
$sl(2,\mathbb{C})$ top is described by the Sklyanin algebra
\cite{Slyanin}. It is the algebra with quadratic relations that
follow from the Yang-Baxter "RLL" equation with the elliptic
Belavin-Drinfeld R-matrix for $sl(2,\mathbb{C})$. The general aim
of this paper is to apply the method of singular gauge
transformations (developed in \cite{Smir2} for the classical
systems) to the quantum case. For example, by means of this
approach we obtain the following rational non-dynamical R-matrix:
\begin{equation}
\label{Rrat} \widetilde{R}^{\,r}(u)=\left[ \begin {array}{cccc}
{\dfrac {u+2\,\eta}{2\,\eta}}&0&0&0
\\\noalign{\medskip}-u \left( u+2\,\eta \right) \gamma&{\dfrac {u}
{2\,\eta}}&1&0\\\noalign{\medskip}-u \left( u+2\,\eta \right)
\gamma&1&{\dfrac {u}{2\,\eta}}&0\\\noalign{\medskip}-u{\gamma}^{2}
\left( u+2\,\eta \right)  \left( 4\,{\eta}^{2}+2\,u\eta+{u}^{2}
\right) &u \left(u+2\,\eta \right) \gamma&u \left( u+2\,\eta
\right) \gamma&{ \dfrac {u+2\,\eta}{2\,\eta}}\end {array} \right]
\end{equation}
This R-matrix satisfies  QYBE for any $\gamma$. In the case of
$\gamma=0$ we get the ordinary 6-vertex rational R-matrix. The
matrix elements of L-operator that satisfies the "RLL" equation
with R-matrix (\ref{Rrat}) obey a quadratic algebra which can be
obtained as a degeneration of Sklyanin algebra. It is well known,
the quantum algebra arising from the "RLL" equation for the
ordinary rational R-matrix (\ref{Rrat}) at $\gamma=0$ is the
Yangian $Y[sl_{2}]$, in this way the rational Sklyanin algebra
arising from $(\ref{Rrat})$ for generic $\gamma$ is the
one-parametric deformation
of $Y[sl_{2}]$.\\
\indent The structure of the article is as follows. In Sect. 2 we
use singular gauge transformations to obtain new trigonometric and
rational solutions of QYBE.  In Sect. 3 we apply this technique to
L-operators. Sect.4 deals with representations of degenerate
Sklyanin algebras by difference operators. In Sect 5 we show that
degenerations of the elliptic R-matrix for $sl(2,\mathbb{C})$,
that we have found, turn out to be related by a quasi-Hopf twist
to the standard degenerations. The $sl(N,\mathbb{C})$-case is
discussed in Sect.6.
\section{New solutions of QYBE via singular gauge transformations }
In this section we discuss briefly the standard trigonometric and
rational degenerations of the elliptic Baxter's R-matrix for
$sl(2,\mathbb{C})$. We give a definition of singular gauge
transformations and use them for obtaining new
degenerations.\\
\\
\indent The elliptic R-matrix for $sl(2,\mathbb{C})$ with two
dimensional auxiliary space has the form:
\begin{equation}
\label{EllR}
R^{\,e}(u)=\sum_{a=0}^{3}W_{a}\left(u+\eta\right)\,\sigma_{a}\otimes\sigma_{a}
\end{equation}
where $W_{a}(u)=W_{a}(u|\eta,\,\tau)$, a=0,...,3 are functions of
a variable $u$ (called spectral parameter) with parameters $\eta$
and $\tau$:
$$
W_{a}\left(u\right)=\dfrac{\theta_{i(a)}\left(u\right)}{\theta_{i(a)}\left(\eta\right)}\,,\
\ \ \ \ i(a)=a+(-1)^{a}.
$$
Here $ \theta_{a}(u)$ are the standard Jacoby theta-functions with
characteristics and the modular parameter $\tau$; $\sigma_{a}$ are
the Pauli matrices ($\sigma_{0}$ is the unit matrix). For
convenience we write out the explicit form of (\ref{EllR}) in
terms of the theta-functions with half integer characteristics:
$$
 R^{\,e}(u)=
 \left[ \begin {array}{cccc} {\frac { \theta_{{1,1}}
 \left( u+\eta,\tau \right) }{ \theta_{{1,1}}   \left(
\eta,\tau \right) }}+{\frac {  \theta_{{1,0}}  \left( u+ \eta,\tau
\right) }{  \theta_{{1,0}}  \left( \eta,\tau
 \right) }}&0&0&{\frac {  \theta_{{0,1}}   \left( u+\eta,
\tau \right) }{  \theta_{{0,1}}   \left( \eta,\tau
 \right) }}-{\frac {  \theta_{{0,0}}   \left( u+\eta,\tau
 \right) }{  \theta_{{0,0}}   \left( \eta,\tau \right) }}
\\\noalign{\medskip}0&{\frac { \theta_{{1,1}}  \left( u
+\eta,\tau \right) }{  \theta_{{1,1}}   \left( \eta,\tau
 \right) }}-{\frac {  \theta_{{1,0}}   \left( u+\eta,\tau
 \right) }{  \theta_{{1,0}}   \left( \eta,\tau \right) }}
&{\frac {  \theta_{{0,1}}   \left( u+\eta,\tau \right) }{
  \theta_{{0,1}}   \left( \eta,\tau \right) }}+{\frac {
  \theta_{{0,0}}   \left( u+\eta,\tau \right) }{
\theta_{{0,0}}   \left( \eta,\tau \right) }}&0
\\\noalign{\medskip}0&{\frac {  \theta_{{0,1}}   \left( u
+\eta,\tau \right) }{  \theta_{{0,1}}  \left( \eta,\tau
 \right) }}+{\frac {  \theta_{{0,0}}  \left( u+\eta,\tau
 \right) }{  \theta_{{0,0}}   \left( \eta,\tau \right) }}
&{\frac { \theta_{{1,1}}   \left( u+\eta,\tau \right) }{
  \theta_{{1,1}}   \left( \eta,\tau \right) }}-{\frac {
  \theta_{{1,0}}  \left( u+\eta,\tau \right) }{
\theta_{{1,0}}   \left( \eta,\tau \right) }}&0
\\\noalign{\medskip}{\frac {  \theta_{{0,1}}   \left( u+
\eta,\tau \right) }{  \theta_{{0,1}}  \left( \eta,\tau
 \right) }}-{\frac {  \theta_{{0,0}}   \left( u+\eta,\tau
 \right) }{  \theta_{{0,0}}  \left( \eta,\tau \right) }}
&0&0&{\frac {  \theta_{{1,1}}  \left( u+\eta,\tau
 \right) }{  \theta_{{1,1}}  \left( \eta,\tau \right) }}
+{\frac {  \theta_{{1,0}}  \left( u+\eta,\tau \right) }{
  \theta_{{1,0}}   \left( \eta,\tau \right) }}
\end {array} \right]
$$
The elliptic R-matrix (\ref{EllR}) satisfies QYBE:
\begin{equation}
\label{QYBE}
 R_{1\,2}^{\,e}(u-v)\,R_{1\,3}^{\,e}(u)\,R_{2\,3}^{\,e}(v)=
R_{2\,3}^{\,e}(v)\,R_{1\,3}^{\,e}(u)\,R_{1\,2}^{\,e}(u-v)
\end{equation}
Here R-matrix $R_{1\,2}(u)$ acts in the tensor product
$\mathbb{C}^{2}\otimes\mathbb{C}^{2}\otimes\mathbb{C}^{2}$ as
$R(u)$ on the first and second spaces, and as the identity
operator on the third one (similarly for $R_{1\,3}$ and
$R_{2,3}$). Solutions of (\ref{QYBE}) play important role  for the
construction of solvable two-dimensional statistical models
\cite{Baxter}, in the integrable systems \cite{Belavin} , and
representation theory \cite{Drinfeld}. Letting the modular
parameter to imaginary infinity $\tau\rightarrow \textrm{i}
\infty$ we get the standard trigonometric R-matrix:
\begin{equation}
\label{RTstan} R^{\,t}(u) =\dfrac{1}{\sin(2\, \pi\, \eta)} \left[
\begin {array}{cccc} \sin \left( \pi \, \left( u+2\,\eta \right)
\right) &0&0&0\\\noalign{\medskip}0&\sin \left( \pi \,u \right)
&\sin \left( 2\,\pi \,\eta \right) &0\\\noalign{\medskip}0& \sin
\left( 2\,\pi \,\eta \right) &\sin \left( \pi \,u \right) &0
\\\noalign{\medskip}0&0&0&\sin \left( \pi \, \left( u+2\,\eta \right)
\right) \end {array} \right]
\end{equation}
To get the standard rational degeneration of (\ref{RTstan}), let
us substitute in (\ref{RTstan}) the constant $\pi$ by a formal
variable $x$. In the limit $x=0$ we get:
\begin{equation}
\label{RRstand}
 R^{\,r}(u)=\dfrac{1}{2\,\eta} \left[ \begin {array}{cccc}
u+2\,\eta&0&0&0\\\noalign{\medskip}0&u&2
\,\eta&0\\\noalign{\medskip}0&2\,\eta&u&0\\\noalign{\medskip}0&0&0&u+2
\,\eta\end {array} \right]
\end{equation}
On the set of solutions of QYBE for $sl(2,\mathbb{C})$ acts the
group $SL(2,\mathbb{C})$ (the gauge group). More precisely, let
$G$ be two-dimensional matrix representing an element from
$SL(2,\mathbb{C})$ in the fundamental representation. We say that
R-matrix $R_{2}$ is the gauge transformation of $R_{1}$ by matrix
$G$ if:
$$
R_{2}= A\,R_{1}\,A^{-1},\ \ \ \ A=G_{1}\,G_{2},\ \ \
G_{1}=G\,\otimes\,1,\ \ \\G_{2}=1\,\otimes\,G
$$
The matrices related by a gauge transformation are called gauge
equivalent.\\
\indent Let us consider the gauge transformation depending on
modular parameter $q=\exp(2\pi\,i\,\tau)$ with the matrix:
\begin{equation}
\label{TrigTrans} G^{\,t}=\left[ \begin {array}{cc}
q^{1/8}\,\alpha^{-1/2}&0
\\\noalign{\medskip}0& \alpha^{1/2}\,q^{-1/8}\end {array}
\right]
\end{equation}
If we take the trigonometric limit $q=0$ after applying gauge
transformation (\ref{TrigTrans}) to  elliptic R-matrix
(\ref{EllR}) we get the following result:
\begin{equation}
\label{TrigR}
 \widetilde{R}^{\,t}(u)=\dfrac{1}{\sin(2\, \pi\, \eta)} \,\left[
\begin {array}{cccc} \sin \left( \pi \, \left( u+2\,\eta \right)
\right) &0&0&0\\\noalign{\medskip}0&\sin \left( \pi \,u \right)
&\sin \left( 2\,\pi \,\eta \right) &0\\\noalign{\medskip}0& \sin
\left( 2\,\pi \,\eta \right) &\sin \left( \pi \,u \right) &0
\\\noalign{\medskip}4\,{\alpha}^{2}\sin \left( \pi \,u \right) \sin \left(
2\,\pi \,\eta \right) \sin \left( \pi \, \left( u+2\,\eta \right)
\right) &0&0&\sin \left( \pi \, \left( u+2\,\eta \right)  \right)
\end {array} \right]
\end{equation}
This matrix, obviously, satisfies QYBE because elliptic R-matrix
(\ref{EllR}) gauge transformed by (\ref{TrigTrans}) satisfies QYBE
for any $q$. It can be shown that R-matrix (\ref{TrigR}) is not
gauge equivalent to (\ref{RTstan}), therefore it defines a new
trigonometric solution of QYBE. Moreover, this R-matrix is the
one-parametric generalization of standard one (\ref{RTstan}) with
the parameter $\alpha$. At the point $\alpha=0$ we get the
standard
trigonometric degeneration.\\
\indent We note that the determinant of the matrix $G^{\,t}$ is
equal to one for any value of the parameter $q$, i.e. for any $q$
this matrix belongs to the gauge group $SL(2,\mathbb{C})$. This
matrix, however, is singular at the point $q=0$ and this is why we
obtain a new trigonometric R-matrix that is not gauge equivalent
to the standard trigonometric degeneration (\ref{RTstan}). We call
gauge transformations with these properties \textit{singular}. The
importance of these transformations lies in the fact that, as in
the examples above, they can be used for obtaining new degenerate
solutions of QYBE.\\
\indent Let us consider another example of a singular gauge
transformation with the matrix:
\begin{equation}
\label{RatTrans}
 G^{\,r}= \left[ \begin {array}{cc} x\,\alpha^{1/2}\,\beta^{-1/2}&0\\\noalign{\medskip}2\,
 \alpha^{1/2}\,\beta^{1/2}\,{x}^{-1}&\beta^{1/2}\,\alpha^{-1/2}\,{x}^{-1}\end {array}
\right]
\end{equation}
The determinant of this matrix is equal to one for any value of
the parameter $x$ (i.e. $G^{r}$ belongs to the gauge group), but
at the point $x=0$ $G^{r}$ becomes singular. To get a rational
degeneration of (\ref{TrigR}) we should substitute the constant
$\pi$ by a formal variable $x$. Then, applying the gauge
transformation with matrix (\ref{RatTrans}) to (\ref{TrigR})  and
taking the limit $x=0$ we get the following rational R-matrix:
\begin{equation}
\label{RatR}
 \widetilde{R}^{\,r}(u)=\left[ \begin {array}{cccc} {\dfrac
{u+2\,\eta}{2\,\eta}}&0&0&0
\\\noalign{\medskip}-u \left( u+2\,\eta \right) \beta&{\dfrac {u}
{2\,\eta}}&1&0\\\noalign{\medskip}-u \left( u+2\,\eta \right)
\beta&1&{\dfrac {u}{2\,\eta}}&0\\\noalign{\medskip}-u{\beta}^{2}
\left( u+2\,\eta \right)  \left( 4\,{\eta}^{2}+2\,u\eta+{u}^{2}
\right) &u \left(u+2\,\eta \right) \beta&u \left( u+2\,\eta
\right) \beta&{ \dfrac {u+2\,\eta}{2\,\eta}}\end {array} \right]
\end{equation}
This R-matrix is the one-parametric generalization of standard
rational R-matrix (\ref{RRstand}): it satisfies QYBE for any value
of the parameter $\beta$, and at the point $\beta=0$ coincides
with
(\ref{RRstand}).\\
\indent We should mention that matrices of gauge transformations
(\ref{TrigTrans}), (\ref{RatTrans}) have special dependence on
their parameters $q$ and $x$: after applying this transformations
we get finite limits at $q=0$, $x=0$. It imposes strict conditions
on the matrix elements of (\ref{TrigTrans}) and (\ref{RatTrans}).
\section{Quantum Lax operators}
The universal elliptic L-operator with 2-dimensional auxiliary
space has the form:
\begin{equation}
\label{EllLax1}
 L^{e}\left( u \right)=\sum_{a=0}^{3}\,W_{a}\left( u
\right)\,S_{a}\otimes \sigma_{a}
\end{equation}
Let us write out the explicit expression of this operator in terms
of the elliptic theta-functions with half-integer characteristics:
\begin{equation}
\label{EllLax}
 L^{e}(u)=\left[ \begin {array}{cc} {\dfrac {
\theta_{{1,1}} \left( u \right) }{ 2\,\theta_{{1,1}}  \left( \eta
\right) }}\,S_{{0}}-{\dfrac {  \theta_{{1,0}}
 \left( u \right) }{  2\,\theta_{{1,0}}  \left( \eta
 \right) }}\,S_{{3}}&{\dfrac {  \theta_{{0,1}}
 \left( u \right) }{ 2\, \theta_{{0,1}}   \left( \eta
 \right) }}\,S_{{1}}+{\dfrac {i\theta_{{0,0}}
 \left( u \right) }{ 2\, \theta_{{0,0}}  \left( \eta
 \right) }}\,S_{{2}}\\\noalign{\medskip}{\dfrac {  \theta_{{0,
1}}  \left( u \right) }{ \theta_{{0,1}}
 \left(2\, \eta \right) }}\,S_{{1}}-{\dfrac {i  \theta_{{0,0}}
   \left( u \right) }{ 2\, \theta_{{0,0}}  \left(
\eta \right) }}\,S_{{2}}&{\dfrac { \theta_{{1,1}}
 \left( u \right) }{ 2\,\theta_{{1,1}}  \left( \eta
 \right) }}\,S_{{0}}+{\dfrac { \theta_{{1,0}}
 \left( u \right) }{  2\,\theta_{{1,0}}  \left( \eta
 \right) }}\,S_{{3}}\end {array} \right]
\end{equation}
Here the operators $S_{0}$, $S_{\alpha}$ $\alpha\,=$ $1$, $2$,
$3$, are the generators of Sklyanin algebra \cite{Slyanin}. These
operators obey the conditions under which L-operator
(\ref{EllLax}) satisfies so called "RLL"  equation:
\begin{equation}
\label{RLL}
R^{e}_{{1,2}}\left(u-v\right)\,L^{e}_{1}(u)\,L^{e}_{2}(v)\,=\,L^{e}_{2}(v)\,L^{e}_{1}(u)\,R^{e}_{{1,2}}(u-v)
\end{equation}
where we use the following standard notations:
$L_{1}(u)=L(u)\,\otimes\,1$ and $L_{2}(u)=1\,\otimes\,L(u)$.\\
\indent Our aim is to find the form of the trigonometric and
rational L-operators that satisfy the "RLL" equation for
R-matrices (\ref{TrigR}) and (\ref{RatR}). To find these operators
we should apply gauge transformations (\ref{TrigTrans}) and
(\ref{RatTrans}) to elliptic L-operator (\ref{EllLax}). A gauge
transformation with matrix a $G$ acts on Lax operators as a
conjugation in the "external" space:
\begin{equation}
\label{LaxTrans}
 L^{'}(u)=G\,L(u)\,G^{-1}
\end{equation}
and as a change of the generators $S_{i}$ by new ones $S_{i}^{'}$
in the "internal" space:
\begin{equation}
\label{Reg} \left[ \begin {array}{cc}
S_{{0}}^{'}-S_{{3}}^{'}&S_{{1}}^{'}+iS_{{2}}^{'}\\\noalign{\medskip}S_{{1}}^{'}-iS_{{2}}^{'}&S_{{0}}^{'}+S_{{3}}^{'}\end
{array}
 \right]=G\,\left[ \begin {array}{cc}
S_{{0}}-S_{{3}}&S_{{1}}+iS_{{2}}\\\noalign{\medskip}S_{{1}}-iS_{{2}}&S_{{0}}+S_{{3}}\end
{array}
 \right]\,G^{-1}
\end{equation}
Let us consider how this scheme works in the trigonometric and
rational cases. The change of the generators induced by gauge
transformation (\ref{Reg}) with matrix (\ref{TrigTrans}) has the
form:
$$
\left[ \begin {array}{cc}
S_{{0}}^{t}-S_{{3}}^{t}&S_{{1}}^{t}+iS_{{2}}^{t}\\\noalign{\medskip}S_{{1}}^{t}-iS_{{2}}^{t}&S_{{0}}^{t}+S_{{3}}^{t}\end
{array}
 \right]=
\left[ \begin {array}{cc} q^{1/8}\,\alpha^{-1/2}&0
\\\noalign{\medskip}0& \alpha^{1/2}\,q^{-1/8}\end {array}
\right]
 \,\left[ \begin {array}{cc}
S_{{0}}^{e}-S_{{3}}^{e}&S_{{1}}^{e}+iS_{{2}}^{e}\\\noalign{\medskip}S_{{1}}^{e}-iS_{{2}}^{e}&S_{{0}}^{e}+S_{{3}}^{e}\end
{array}
 \right]\,\left[ \begin {array}{cc}
q^{1/8}\,\alpha^{-1/2}&0
\\\noalign{\medskip}0& \alpha^{1/2}\,q^{-1/8}\end {array}
\right]
 $$
 or more explicitly:
 \begin{equation}
 \label{BasChenTrig}
 \begin{array}{c}
 S_{0}^{\,e}=S_{0}^{\,t},\\
 \\
 S_{3}^{\,e}=S_{3}^{\,t}\\
 \\
 S_{1}^{\,e}=\dfrac{\alpha}{2\,i}\,q^{-1/4}\left(\,S_{1}^{\,t}+i\,S_{2}^{\,t}\,\right)+
 \dfrac{1}{2\,i\,\alpha}\,q^{1/4}\,\left(\,S_{1}^{\,t}-i\,S_{2}^{\,t}
 \right)\\
 \\
 S_{2}^{\,e}=\dfrac{\alpha}{2}\,q^{-1/4}\left(\,S_{1}^{\,t}+i\,S_{2}^{\,t}\,\right)+
 \dfrac{1}{2\,\alpha}\,q^{1/4}\,\left(\,S_{1}^{\,t}-i\,S_{2}^{\,t}
\right)
\end{array}
 \end{equation}
 Substituting these expressions into
 $G^{\,t}\,L^{\,e}(u)\,(G^{\,t})^{-1}$ and taking the limit $q\rightarrow
 0$ we get the following trigonometric Lax operator:
 \begin{equation}
\label{TrigLax}
 L^{t}(u)=\left[ \begin {array}{cc} S_{{0}}^{\,t}\sin \left( \pi \,u \right) \cot
 \left( \pi \,\eta \right) -S_{{1}}^{\,t} \sin
 \left( \pi \,u \right) &  S_{{1}}^{\,t}+iS_{{2}}^{\,t}  \\\noalign{\medskip}{\alpha}^{2}
 \left( S_{{1}}^{\,t}+iS_{{ 2}}^{\,t} \right)
\left( \cos \left( 2\,\pi \,u \right) -\cos \left( 2\, \pi \,\eta
\right)  \right) +
  S_{{1}}^{\,t}-iS_{{2}}^{\,t}  &S_
{{0}}^{\,t}\sin \left( \pi \,u \right) \cot \left( \pi \,\eta
\right) +S_{{1 }}^{\,t} \sin \left( \pi \,u \right)
\end {array} \right]
\end{equation}
This Lax operator is a one-parametric deformation of the standard
trigonometric degeneration of (\ref{EllLax}). It coincides with
the
standard trigonometric L-operator at the point $\alpha=0$.\\
\indent Let us move on to the rational case. The change of the
generators of the algebra induced by gauge transformation
(\ref{RatTrans}) has the form:
$$
\left[ \begin {array}{cc}
S_{{0}}^{r}-S_{{3}}^{r}&S_{{1}}^{r}+iS_{{2}}^{r}\\\noalign{\medskip}S_{{1}}^{r}-iS_{{2}}^{r}&S_{{0}}^{r}+S_{{3}}^{r}\end
{array}
 \right]=
$$
$$
\left[ \begin {array}{cc}
x\,\alpha^{1/2}\,\gamma^{-1/2}&0\\\noalign{\medskip}2\,\alpha^{1/2}\,\beta^{1/2}\,{x}^{-1}&\beta^{1/2}\,\alpha^{-1/2}\,{x}^{-1}\end
{array} \right]\,\left[ \begin {array}{cc}
S_{{0}}^{t}-S_{{3}}^{t}&S_{{1}}^{t}+iS_{{2}}^{t}\\\noalign{\medskip}S_{{1}}^{t}-iS_{{2}}^{t}&S_{{0}}^{t}+S_{{3}}^{t}\end
{array}
 \right]\,\left[ \begin {array}{cc} x\,\alpha^{1/2}\,\beta^{-1/2}&0\\\noalign{\medskip}2\,\alpha^{1/2}\,\beta^{1/2}\,{x}^{-1}&
 \beta^{1/2}\,\alpha^{-1/2}\,{x}^{-1}\end {array}
\right]^{-1}.
$$
Or more explicitly:
\begin{equation}
\label{BasChenRat}
\begin{array}{c}
S^{\,t}_{0}=S^{\,r}_{0}\\
\\
S_{1}^{\,t}=-{\dfrac { \left(
-{\beta}^{2}-{x}^{4}{\alpha}^{2}+4\,{\beta}^{2}{\alpha}^ {2}
\right) S_{{1}}^{\,r}}{2\,\beta\,{x}^{2}\alpha}}-\,{\dfrac {
\left( -i{\beta
}^{2}+i{x}^{4}{\alpha}^{2}+4\,i{\beta}^{2}{\alpha}^{2} \right)
S_{{2}}^{\,r}}{2\,\beta \,{x}^{2}\alpha}}+2\,\alpha\,S_{{3}}^{\,r}\\
\\
S_{2}^{\,t}={\dfrac {-1/2\,i \left(
{\beta}^{2}-{x}^{4}{\alpha}^{2}+4\,{\beta}^{2}{\alpha}^ {2}
\right) S_{{1}}^{\,r}}{\beta\,{x}^{2}\alpha}}-{\dfrac {1/2\,i
\left( i{\beta
}^{2}+i{x}^{4}{\alpha}^{2}+4\,i{\beta}^{2}{\alpha}^{2} \right)
S_{{2}}^{\,r}}{ \beta\,{x}^{2}\alpha}}+2\,i\,\alpha\,S_{{3}}^{\,r}\\
\\
S_{3}^{\,t}=-2\,{\dfrac {S_{{1}}^{\,r}\beta}{{x}^{2}}}-{\dfrac
{2\,i\,S_{{2}}^{\,r}\beta}{{x}^{2 }}}+S_{{3}}^{r}
\end{array}
\end{equation}
To get the rational limit, in complete analogy with the previous
section, we should substitute in (\ref{TrigLax}) the constant
$\pi$ by a formal variable $x$. Then, substituting
(\ref{BasChenRat}) into $G^{\,r}\,L^{t}(u)\,(G^{\,r})^{-1}$ and
taking the limit $x=0$ we get the following rational L-operator:
$$
L^{\,r}(u)=\left[ \begin {array}{cc}
L_{{1,1}}^{r}&L_{{1,2}}^{r}\\\noalign{\medskip}L_{{2,1}}^{r}&L_{{2,2}}^{r}\end
{array} \right]
$$
where:
$$
L^{\,r}_{{1,1}}=\dfrac{1}{2\,\eta}\, \left(
-S_{{1}}^{\,r}\eta\,{u}^{2}-i\,S_{{2}}^{\,r}{u}^{2}\eta+i\,S_{{2}
}^{\,r}{\eta}^{3}+S_{{1}}^{\,r}{\eta}^{3} \right)
\beta-\dfrac{1}{2\,\eta}\,\left(S_{
{3}}^{\,r}\eta-S_{{0}}^{\,r}u\right)
$$
$$
L^{\,r}_{{1,2}}=1/2\,S_{{1}}^{r}+1/2\,i\,S_{{2}}^{r}
$$
$$
L^{\,r}_{{2,1}}=\left(
3/2\,i{\eta}^{4}S_{{2}}^{r}-i{\eta}^{2}{u}^{2}S_{{2}}^{r}-1/2\,i{u}^{4}S_{{2}}^{r}
-1/2\,{u}^{4}S_{{1}}^{\,r}-{u}^{2}S_{{1}}^{\,r}
{\eta}^{2}+3/2\,S_{{1}}^{\,r}{ \eta}^{4} \right) {\beta}^{2}+
\left( S_{{2}}^{\,r}{u}^{2}-S_{{2}}^{\,r}{\eta}^{2 } \right)
\beta-1/2\,iS_{{2}}^{\,r}+1/2\,S_{{1}}^{\,r}
$$
$$
L_{{2,2}}^{\,r}=1/2\,{\frac { \left(
-S_{{1}}^{\,r}{\eta}^{3}+i\,S_{{2}}^{\,r}{u}^{2}\eta-i\,S_{{2}}^{\,r}{
\eta}^{3}+S_{{1}}^{\,r}{u}^{2}\eta \right)
\beta}{\eta}}+1/2\,{\frac {S_{{3
}}^{\,r}\eta+S_{{0}}^{\,r}u}{\eta}}
$$
Let us introduce the trigonometric and rational Sklyanin algebras
as algebras with four generators $S_{i}^{t}$ and $S_{i}^{r}$
obeying relations following from the "RLL" equations:
\begin{equation}
\label{RLLt}
R^{t}_{{1,2}}\left(u-v\right)\,L^{t}_{1}(u)\,L^{t}_{2}(v)\,=\,L^{t}_{2}(v)\,L^{t}_{1}(u)\,R^{t}_{{1,2}}(u-v)
\end{equation}
\begin{equation}
\label{RLLr}
R^{r}_{{1,2}}\left(u-v\right)\,L^{r}_{1}(u)\,L^{r}_{2}(v)\,=\,L^{r}_{2}(v)\,L^{r}_{1}(u)\,R^{r}_{{1,2}}(u-v)
\end{equation}
In sections 4 and 5 we study these algebras. We write out the
relations for $S_{i}^{t}$ and $S_{i}^{r}$ explicitly, and find
their representations in terms of the difference operators acting
on the space of meromorphic functions on one variable.

\section{Degenerate Slkyanin algebras}
Sklyanin algebra \cite{Slyanin} is the algebra with  four
generators $S_{i}^{\,e}$, $i=0,...,3$ and the following quadratic
relations:
\begin{equation}
\label{EllRel}
[S_{i}^{\,e},S_{j}^{\,e}]_{-}=i\,[S_{0}^{\,e},S_{k}^{\,e}]_{+},\ \
\ \
[S_{0}^{\,e},S_{k}^{\,e}]_{-}=i\,J_{i,j}\,[S_{i}^{\,e},S_{j}^{\,e}]_{+},\
\ \ J_{i,j}=\dfrac{J_{j}-J_{i}}{J_{k}};
\end{equation}

where $ [A,B]_{\pm}=A\,B \pm B\,A $. A triple of indices $(i,j,k)$
in (\ref{EllRel}) stands for any cyclic permutation of $(1,2,3)$.
Structure constants $J_{i}$ have the form:
\begin{equation}
J_{1}={\frac {  \theta_{{0,1}}   \left( 2\,\eta \right)
  \theta_{{0,1}}   \left( 0 \right) }{
\theta_{{0,1}}   \left( \eta \right)^{2}}}\ \ \ \ J_{2}={\frac {
\theta_{{0,0}}  \left( 2\,\eta \right)
  \theta_{{0,0}}   \left( 0 \right) }{
\theta_{{0,0}}  \left( \eta \right)^{2}}}\ \ \ \ J_{3}={\frac {
\theta_{{1,0}}  \left( 2\,\eta \right)
  \theta_{{1,0}}  \left( 0 \right) }{
\theta_{{1,0}} \left( \eta \right)^{2}}}
\end{equation}
Relations (\ref{EllRel}) was introduced by E.Slkyanin as the
minimal set of conditions under which L-operator (\ref{EllLax1})
satisfies the "RLL" equation with elliptic R-matrix (\ref{EllR}).
This algebra possesses two independent central elements (the
Casimir elements):
\begin{equation}
\label{EllCas}
\Omega_{1}^{\,e}={S_{0}^{\,e}}^{2}+{S_{1}^{\,e}}^{2}+{S_{2}^{\,e}}^{2}+{S_{3}^{\,e}}^{2}\,
,\ \ \
\Omega_{2}^{\,e}=J_{1}\,{S_{1}^{\,e}}^{2}+J_{2}\,{S_{2}^{\,e}}^{2}+J_{3}\,{S_{3}^{\,e}}^{2}
\end{equation}
In \cite{Slyanin} the following representation of Sklyanin algebra
in terms of the difference operators was found:
\begin{equation}
\label{EllRepres}
\begin{array}{c}
S_{i}^{\,e}=\dfrac{\chi_{i}(u-s\,\eta)}{\theta_{{1,1}}   \left(
2\,u
\right)}\,\exp(\eta\,\partial_{u})-\dfrac{\chi_{i}(-u-s\,\eta)}{\theta_{{1,1}}
\left( 2\,u \right)}\,\exp(-\eta\,\partial_{u})\\
\\
\chi_{0}=\theta_{{1,1}}   \left( 2\,u
\right)\,\theta_{{1,1}}\left( \eta \right),\ \
\chi_{1}=\theta_{{0,1}}   \left( 2\,u
\right)\,\theta_{{0,1}}\left( \eta \right),\ \
\chi_{2}=\theta_{{0,0}}   \left( 2\,u
\right)\,\theta_{{0,0}}\left( \eta \right),\ \
\chi_{3}=\theta_{{1,0}}   \left( 2\,u \right)\,\theta_{{1,0}}
\left( \eta \right)
\end{array}
\end{equation}
with standard notations $\exp(\pm \eta\,\partial_{u})(f)\left( u
\right)=f\left( u \pm \eta\right)$. In this representations
central elements (\ref{EllCas}) are given by scalar operators:
\begin{equation}
\Omega_{1}^{\,e}=4\,\theta_{{1,1}}^{2}\left( \,(2\,s+1)\,\eta
\,\right),\ \ \ \ \ \Omega_{2}^{\,e}=4\,\theta_{{1,1}}\left(\,
2\,(s+1)\,\eta \,\right)\,\theta_{{1,1}}\left(\, 2\,\eta
\,\right).
\end{equation}
Let us consider the degenerations of the Sklyanin algebra related
to trigonometric and rational R-matrices (\ref{TrigR}) and
(\ref{RatR}). Gauge transformation (\ref{TrigTrans}), as shown
above, induces the transformation for the generators of Sklyanin
algebra (\ref{BasChenTrig}). Substituting these expressions into
quadratic relations (\ref{EllRel}) and expanding them in powers of
$q$ up to the first non-trivial order, we have:
\begin{equation}
\label{TrigRel}
\begin{array}{c}
[\,S_{i}^{\,t},\,S_{j}^{\,t}\,]_{-}=i\,[\,S_{0}^{\,t},\,S_{k}^{\,t}\,]_{+},\\

\\

[\,S_{0}^{\,t},\,S_{1}^{\,t}\,]_{-}=\dfrac{C_{2}}{4}\,[\,S_{1}^{\,t},\,S_{3}^{\,t}\,]_{+}+
\dfrac{i\,(\,2\,C_{1}-C_{2})}{4}\,[\,S_{2}^{\,t},\,S_{3}^{\,t}\,]_{+}\\

\\

[\,S_{0}^{\,t},\,S_{2}^{\,t}\,]_{-}=-\dfrac{i\,(C_{2}+2\,C_{1})}{4}\,[\,S_{1}^{\,t},\,S_{3}^{\,t}\,]_{+}-
\dfrac{C_{2}}{4}\,[\,S_{2}^{\,t},\,S_{3}^{\,t}]_{+}\\

\\

[\,S_{0}^{\,t},\,S_{3}^{\,t}\,]_{-}=\dfrac{C_{3}}{2}\left(\,
{S_{1}^{\,t}}^{2}-{S_{2}^{\,t}}^{2}
-i\,[\,S_{1}^{\,t},\,S_{2}^{\,t}\,]_{+}\, \right)
\end{array}
\end{equation}
where the constants $C_{1}$, $C_{2}$ and $C_{3}$ have the form:
\begin{eqnarray}
 &&C_{1}= \lim_{q\rightarrow 0} \left(J_{{3,1}}-J_{{1,2}}\right)=-2\,{\frac {
\sin \left( \pi \,\eta \right)  ^{2}}{
  \cos \left( \pi \,\eta \right)^{2}}}\\
  &&C_{2}=\lim_{q\rightarrow 0} \left( J_{{1,2}}+J_{{3,1}} \right)=-16\,{\frac {  \sin \left( \pi \,\eta \right)^{2}\cos
 \left( 2\,\pi \,\eta \right) }{ \cos \left( \pi \,\eta
 \right)^{2}}}\\
 &&C_{3}=\lim_{q \rightarrow 0} \dfrac{J_{2,3}}{q}=4\,\sin \left( 2\,\pi \,\eta \right)^{2}
\end{eqnarray}
The algebra generated by four elements $S_{i}^{\,t}$, $i=0,...,3$
with relations (\ref{TrigRel}) represent a natural trigonometric
degeneration of the Sklyanin algebra, and we call it
\textit{trigonometric Sklyanin algebra.} This algebra possesses
two independent Casimirs:
\begin{equation}
\label{TrigCas}
 \Omega_{1}^{\,t}={S_{0}^{\,t}}^2+{S_{1}^{\,t}}^2+{S_{2}^{\,t}}^2+{S_{3}^{\,t}}^2\, ,\ \ \
\Omega_{2}^{\,t}=-{S_{1}^{\,t}}^2-{S_{2}^{\,t}}^{2}-L_{1}\,{S_{3}^{\,t}}^2-\dfrac{L_{3}}{4}\,\left({S_{1}^{\,t}}^2-{S_{2}^{\,t}}^2
-i\,[S_{1}^{\,t},S_{2}^{\,t}]_{+} \right)
\end{equation}
with the following constants:
$$
L_{1}={\frac {\cos \left( 2\,\pi \,\eta \right) }{  \cos \left(
\pi \, \eta \right) ^{2}}},\ \ \ \ L_{3}=48\, \cos \left( \pi
\,\eta \right) ^{2}-16-32\,
 \cos \left( \pi \,\eta \right)^{4}
$$
Substituting the generators (\ref{EllRepres}) into
(\ref{BasChenTrig}) and taking the limit $q=0$ we find the
following representations for the generators of the trigonometric
Sklyanin algebra:

$$
\begin{array}{c}
 S_{0}^{\,t}={\dfrac { -\cos \left( \pi \, \left(
2\,u-\eta-2\,s\eta \right)
 \right) +\cos \left( \pi \, \left( 2\,u+\eta-2\,s\eta \right)
 \right)}{\sin \left( 2\,\pi \,u
 \right) }}\exp(\eta\,\partial_{u})+\\
 {\dfrac {  \cos \left( \pi \, \left( \eta+2\,u+2\,s
\eta \right)  \right) -\cos \left( \pi \, \left(
-\eta+2\,u+2\,s\eta
 \right)  \right)}{\sin \left( 2\, \pi \,u \right)
 }}\exp(-\eta\,\partial_{u})\\
 \\
S_{1}^{\,t}={\dfrac { 1-2\,\cos \left( 2\,\pi \,\eta \right)
-2\,\cos
 \left( 4\,\pi \, \left( u+s\eta \right)  \right)  }{2\,\sin \left( 2\,\pi \,u \right)
 }}\exp(-\eta\,\partial_{u})-\\
{\dfrac {  1-2\,\cos \left( 2\,\pi \,\eta \right) -2\,\cos \left(
4\,\pi \,
 \left( -u+s\eta \right)  \right)
 }{2\,
\sin \left( 2\,\pi \,u \right) }}\exp(\eta\,\partial_{u})\\
\\
S_{2}^{\,t}={\dfrac {i \left( 2\,\cos \left( 2\,\pi \,\eta \right)
+2\,\cos
 \left( 4\,\pi \, \left( u+s\eta \right)  \right) +1 \right)  }{2\,\sin \left( 2\,\pi \,u \right)
 }}\exp(-\eta\,\partial_{u})-\\
{\dfrac {i
 \left( 2\,\cos \left( 2\,\pi \,\eta \right) +2\,\cos \left( 4\,\pi
\, \left( -u+s\eta \right)  \right) +1 \right)}{2\,\sin \left(
2\,\pi \,u \right) }}\exp(\eta\,\partial_{u})\\
\\
 S_{3}^{\,t}=-{\dfrac {
\cos \left( \pi \, \left( -\eta+2\,u+2\,s\eta
 \right)  \right) +\cos \left( \pi \, \left( \eta+2\,u+2\,s\eta
 \right)  \right) }{\sin \left( 2\,
\pi \,u \right) }}\exp(-\eta\,\partial_{u}) +\\
 {\dfrac {  \cos
\left( \pi \, \left( -\eta-2\, u+2\,s\eta \right) \right) +\cos
\left( \pi \, \left( \eta-2\,u+2\,s \eta \right) \right) }{\sin
\left( 2 \,\pi \,u \right) }}\exp(\eta\,\partial_{u})
\end{array}
$$
Substituting these difference operators into (\ref{TrigCas}) we
find that in this representation the Casimir operators have the
values:
\begin{equation}
\Omega_{1}^{\,t}=16\, \left( \sin \left( \eta\,\pi \, \left(
1+2\,s \right)  \right)\,
 \right) ^{2},\ \ \ \
 \Omega_{2}^{\,t}=16\,\sin \left( 2\,\pi \,\eta\, \left( s+1 \right)  \right) \sin
 \left( 2\,\pi \,\eta\,s \right)
\end{equation}
We should note here, that the trigonometric degeneration for the
Sklyanin algebra of this kind appeared for the first time in
\cite{GorskyZabrodin}. Our approach differs from
\cite{GorskyZabrodin} by different choosing of the generators for
the degenerate algebra. \\
\indent To find a rational degeneration we note that gauge
transformation (\ref{RatTrans}) induces the transformation for the
generators of trigonometric Sklyanin algebra (\ref{BasChenRat}).
Substituting (\ref{BasChenRat}) to the relations (\ref{TrigRel})
and taking the limit $x=0$ (we replace everywhere the constant
$\pi$ by a formal parameter $x$ ) we get the following relations
for the generators of  \textit{the rational Sklyanin algebra}:
\begin{equation}
\begin{array}{c}
[S_{1}^{\,r},S_{2}^{\,r}]_{-}=i\,[S_{0}^{\,r},S_{3}^{\,r}]_{+},\ \
\ \
[S_{2}^{\,r},S_{3}^{\,r}]_{-}=i\,[S_{0}^{\,r},S_{1}^{\,r}]_{+},\ \
\ \
[S_{3}^{\,r},S_{1}^{\,r}]_{-}=i\,[S_{0}^{\,r},S_{2}^{\,r}]_{+}\\

\\

[\,S_{0}^{\,r},\,S_{3}^{\,r},]_{-}=16\,i\,\eta^4\,[S_{1}^{\,r},S_{3}^{\,r}]_{+}-2\,i\,\eta^2[S_{2}^{\,r},S_{3}^{\,r}]_{+}-16\,\eta^4\,\left(\,
{S_{1}^{\,r}}^2-{S_{2}^{\,r}}^2 \right)\\

\\

[S_{0}^{\,r},S_{2}^{\,r}]_{-}=-8\,\eta^4\,[S_{2}^{\,r},S_{3}^{\,r}]_{+}+2\,\left(8\,\eta^4-1
\right)\,\eta^2\,[S_{1}^{\,r},S_{2}^{\,r}]_{+} -
8\,i\,\eta^4\,[S_{1}^{\,r},S_{3}^{\,r}]_{+}\\

\\

+4\,i\,\eta^2\,\left(4\,\eta^4-1\right)\,{S_{1}^{\,r}}^{2}-16\,i\,\eta^6\,{S_{2}^{\,r}}^2
+4\,i\,\eta^2\,{S_{3}^{\,r}}^{2}\\

\\

[S_{0}^{\,r},S_{1}^{\,r}]_{-}=-8\,i\,\eta^4\,[S_{2}^{\,r},S_{3}^{\,r}]_{+}+2\,i\,\left(
8\,\eta^4+1
\right)\,\eta^2\,[S_{1}^{\,r},S_{2}^{\,r}]_{+}+8\,\eta^2\,[S_{1}^{\,r},S_{3}^{\,r}]_{+}\\

\\

-16\,\eta^6\,{S_{1}^{\,r}}^{2}+4\,\left(4\,\eta^4+1
\right)\,\eta^2\,{S_{2}^{\,r}}^{2}-4\,\eta^{2}\,{S_{3}^{\,r}}^2
\end{array}
\end{equation}
The central elements of this algebra have the form:
\begin{equation}
\begin{array}{c}
\Omega_{1}^{\,r}={S_{0}^{\,r}}^2+{S_{1}^{\,r}}^2+{S_{2}^{\,r}}^2+{S_{3}^{\,r}}^2,\\
\\
\Omega_{2}^{\,r}=\left( 1-12\,\eta^4
\right)\,{S_{1}^{\,r}}^{2}+\left(12\,\eta^4+1
\right)\,{S_{2}^{\,r}}^2+{S_{3}^{\,r}}^{2}+12\,i\,\eta^4\,[S_{1}^{\,r},S_{2}^{\,r}]_{+}+2\,\eta^2\,[S_{1}^{\,r},S_{3}^{\,r}]_{+}
-2\,i\,\eta^2\,[S_{2}^{\,r},S_{3}^{\,r}]_{+}
\end{array}
\end{equation}
As well as in the trigonometric case we find the difference
operators, which represent the generators of this algebra:
$$
S_{0}^{\,r}={\frac { \left( -2\,u\eta-2\,s{\eta}^{2} \right)
}{u}}\exp(-\eta\,\partial_{u})+{\frac { \left(
-2\,u\eta+2\,s{\eta}^{2} \right)  }{u}}\exp(\eta\,\partial_{u})
$$
$$
S_{1}^{\,r}={\frac { \left(
8\,{s}^{2}{\eta}^{4}-64\,{u}^{3}s\eta+16\,us{\eta
}^{3}-64\,u{s}^{3}{\eta}^{3}-96\,{u}^{2}{s}^{2}{\eta}^{2}-{\eta}^{4}+8
\,{u}^{2}{\eta}^{2}-16\,{u}^{4}-16\,{s}^{4}{\eta}^{4}+1 \right)
}{4\,u}}\exp(-\eta\,\partial_{u})
$$

$$
 +{\frac { \left( 16\,{s}^{4}{\eta}^{4
}+{\eta}^{4}-64\,{u}^{3}s\eta+16\,{u}^{4}-64\,u{s}^{3}{\eta}^{3}+96\,{
u}^{2}{s}^{2}{\eta}^{2}-1-8\,{s}^{2}{\eta}^{4}-8\,{u}^{2}{\eta}^{2}+16
\,us{\eta}^{3} \right) }{4\,u}}\exp(\eta\,\partial_{u})
$$
$$
S_{2}^{\,r}={\frac { \left(
-8\,i{s}^{2}{\eta}^{4}+i+64\,i{s}^{3}{\eta}^{3}u-
16\,is{\eta}^{3}u+16\,i{s}^{4}{\eta}^{4}+i{\eta}^{4}+96\,i{u}^{2}{s}^{
2}{\eta}^{2}-8\,i{u}^{2}{\eta}^{2}+64\,i{u}^{3}s\eta+16\,i{u}^{4}
 \right)}{4\,u}}\exp(-\eta\,\partial_{u})
$$
$$
+\frac { \left( 64\,i{s}^{
3}{\eta}^{3}u+8\,i{u}^{2}{\eta}^{2}-96\,i{u}^{2}{s}^{2}{\eta}^{2}-16\,
is{\eta}^{3}u+8\,i{s}^{2}{\eta}^{4}-i{\eta}^{4}+64\,i{u}^{3}s\eta-16\,
i{s}^{4}{\eta}^{4}-i-16\,i{u}^{4} \right)
}{4\,u}\exp(\eta\,\partial_{u})
$$
$$
S_{3}^{\,r}={\frac { \left(
{\eta}^{2}+4\,{u}^{2}+8\,us\eta+4\,{s}^{2}{\eta}^ {2} \right)
}{2\,u}}\exp(-\eta\,\partial_{u})+{\frac { \left( -4\,{u}
^{2}+8\,us\eta-{\eta}^{2}-4\,{s}^{2}{\eta}^{2} \right)
}{u}}\exp(\eta\,\partial_{u})
$$
Substituting these generators into the expressions for the central
elements, we find that in this representation they acts as scalar
operators with the following eigenvalues:
\begin{equation}
\Omega_{1}^{\,r}=16\,{\eta}^{2}  \left( 2\,s+1 \right) ^{2}, \ \ \
\ \Omega_{2}^{\,r}=64\,{\eta}^{2}s\,  \left( 1+s \right)
\end{equation}
\section{Quasi-Hopf twist}
In this section we demonstrate that standard trigonometric and
rational R-matrices (\ref{RTstan}) and (\ref{RRstand}) related by
the quasi-Hopf twist to non-standard degenerations (\ref{TrigR})
and (\ref{RatR}). To show this, we construct the matrices of
quasi-Hopf twists for these cases "by hands".\\
\indent We start from the trigonometric case. Let us consider the
operator acting in $\mathbb{C}^{2}\otimes\mathbb{C}^{2}$, with the
matrix:
\begin{equation}
\label{Qt} Q^{t}(u)= \left[ \begin {array}{cccc}
1&0&0&0\\\noalign{\medskip}0&1&0&0
\\\noalign{\medskip}0&0&1&0\\\noalign{\medskip}2\,{\alpha}^{2}\sin \left(
\pi \,u \right) \sin \left( 2\,\pi \,\eta \right) &0&0&1\end
{array}
 \right]
\end{equation}
This operator has the following properties:
\begin{equation}
\label{propt} Q^{t}(u)^{-1}=Q^{t}(-u),\ \ \
P\,Q^{t}(u)\,P=Q^{t}(u),
\end{equation}
where $P$ is the permutation operator in
$\mathbb{C}^{2}\otimes\mathbb{C}^{2}$.  We find that trigonometric
R-matrices (\ref{TrigR}) and (\ref{RTstan}) related by this
operator in the following way:
\begin{equation}
\widetilde{R}^{t}(u)=Q^{t}(u)\,R^{t}(u)\,Q^{t}(u)
\end{equation}
Using (\ref{propt}) we can rewrite this relation in the form of a
quasi-Hopf twist:
\begin{equation}
\label{twistT} \widetilde{R}^{t}(u-v)\,F^{t}_{2\,1}(v,u)=
F^{t}_{1\,2}(u,v)\,R^{t}(u-v),
\end{equation}
where
$$
F_{1\,2}^{t}(u,v)=Q^{t}(u-v),\ \ \
F_{2\,1}^{t}(u,v)=P\,F^{t}_{1\,2}(u,v)\,P
$$
Similarly, in the rational case, we find the following matrix:
\begin{equation}
\label{Qr} Q^{r}(u)=\left[ \begin {array}{cccc}
1&0&0&0\\\noalign{\medskip}-u\,\beta\,\eta&1&0&0\\\noalign{\medskip}-u\,\beta\,\eta&0&1&0\\\noalign{\medskip}-
\eta{u}^{3}\,{\beta}^{2}-{\eta}
^{2}\,{u}^{2}\,{\beta}^{2}-4\,{\eta}^{3}\,u\,{\beta}^{2}&u\,\beta\,h&u\,\beta\,h&1\end
{array} \right].
\end{equation}
This operator has the following properties:
\begin{equation}
\label{propr} Q^{r}(u)^{-1}=Q^{r}(-u),\ \ \
P\,Q^{r}(u)\,P=Q^{r}(u)
\end{equation}
Using this matrix, we can relate rational R-matrices (\ref{RatR})
and (\ref{RRstand}) in the following way:
\begin{equation}
\label{rel} \widetilde{R}^{r}(u)=Q^{r}(u)\,R^{r}(u)\,Q^{r}(u),
\end{equation}
Similarly to the trigonometric case, using (\ref{propr}) we can
rewrite this equation in the form of a quasi-Hopf twist:
\begin{equation}
\label{twistR} \widetilde{R}^{\,r}(u-v)\,F^{\,r}_{2\,1}(v,u)=
F^{\,r}_{1\,2}(u,v)\,R^{\,r}(u-v),
\end{equation}
where
$$
F_{1\,2}^{\,r}(u,v)=Q^{\,r}(u-v),\ \ \
F_{2\,1}^{\,r}(u,v)=P\,F^{\,r}_{1\,2}(u,v)\,P.
$$
Therefore, equations (\ref{twistT}) and (\ref{twistR}) show that
the standard and nonstandard degenerate R-matrices turn out to be
related by the quasi-Hopf twists $F^{\,t}_{1\,2}$ and
$F^{\,r}_{1\,2}$.
\section{$ sl(N,\mathbb{C}) $-case}

In this section we obtain non-standard trigonometric and rational
degenerations of the elliptic Belavin's R-matrix for
$sl(N,\mathbb{C})$. The elliptic R-matrix for $sl(N,\mathbb{C})$
has the following form:
\begin{equation}
\label{ELLRforSLN} R^{e}(u)=\sum\limits_{i^{'},\,
j^{'}=1}^{N}\delta_{|i+j||i^{'}+j^{'}|}\dfrac{\theta^{(i^{'}-j^{'})}(u+2\,\eta)\,\theta^{(0)}(u)}
{\theta^{(i^{'}-i)}(2\,\eta)\,\theta^{(i-j^{'})}(u)}\,E_{i\,i^{'}}\otimes
E_{j\,j^{'}}
\end{equation}
where $(E_{i\,i^{'}})_{k\,l}=\delta_{i\,k}\,\delta_{i^{'}l}$ are
elements of standard basis in the space of $N^2\times N^{2}$
matrices, $|i|=i\mod N$, and we use the following notations for
the theta-functions:
\begin{equation}
\theta^{(j)}(u)=\sum_{m\in\,\mathbb{Z}}\exp\left(
\pi\,i\,N\,\tau(m+\dfrac{1}{2}-\dfrac{j}{N})^2+2\,\pi\,i\,(m+\dfrac{1}{2}
-\dfrac{j}{N})(u+\dfrac{1}{2}) \right)
\end{equation}
Similarly to the $N=2$ case, to get the  non-standard
degenerations, we should apply the additional gauge
transformations before taking the corresponding limits. Given the
matrices of these transformations $G^{t}(q)$ and $G^{r}(x)$, we
can find the nonstandard degenerations by calculating the
following limits:
\begin{equation}
\label{TrigLim} \widetilde{R}^{t}(u)=\lim_{q\rightarrow
0}G^{t}_{1}(q)\,G_{2}^{t}(q)\,R^{e}(u)\,(G^{t}_{1}(q)\,G_{2}^{t}(q))^{-1}
\end{equation}
\begin{equation}
\label{RatLim} \widetilde{R}^{r}(u)=\lim_{x\rightarrow
0}G^{r}_{1}(x)\,G_{2}^{r}(x)\,R^{t}(u)\,(G^{r}_{1}(x)\,G_{2}^{r}(x))^{-1}
\end{equation}
The main difficulty here is to find the explicit form for the
matrix elements of $G^{t}(q)$ and $G^{r}(x)$. This problem was
solved in our previous work \cite{Smir2}. We found that these
matrices have the following structure:
\begin{equation}
\label{Gtrig}
(G^{t}(q))_{i,\,j}=\delta_{i,\,j}\,q^{-(\frac{N}{2}(\frac{i^2}{N^2}-\frac{i}{n})+\frac{N^2-1}{12\,N})}
\end{equation}
The matrix for rational transforation $G^{r}(x)$ can be presented
as a product:
\begin{equation}
\label{Grat} G^{r}(x)=S_{2}\,S_{1}
\end{equation}
where $S_{2}$ is a diagonal matrix:
\begin{equation}
S_{2}=\textrm{diag}(x^{b_{1}},x^{b_{2}},...,x^{b_{N-1}},x^{b_{N}}),\
\ \ b_{i}=-\frac{N(N-1)}{2N}+\frac{1-(i-1)}{N}
\end{equation}
and $S_{2}$ is a constant matrix (i.e. is independent of $x$):
\begin{equation}
(S_{2})_{i,j}=\varepsilon(i\neq
N)\varepsilon(j\leq\,i)\dfrac{(i-1)!}{(j-i)!(j-1)!}+\varepsilon(i=N)\dfrac{N!}{j!(N-j)!}
\end{equation}
 Here $\varepsilon$(condition) is equal to 1 if the condition is
true and 0 otherwise. Knowing matrices of gauge transformations
(\ref{Gtrig}) and (\ref{Grat}) we can easily use modern programs
such as Maple or Mathematica and formulas (\ref{TrigLim}) and
(\ref{RatLim}) to compute the nonstandard trigonometric and
rational R-matrices for $N\leq 15$ . Here, as an example, we
present the results of these calculations for N=3. For the
$sl(3,\mathbb{C})$ non-standard trigonometric R-matrix we find:
\begin{equation}
\widetilde{R}^{trig}(u)=\left[ \begin {array}{ccc}
R_{{1,1}}^{t}&R_{{1,2}}^{t}&R_{{1,3}}^{t}\\\noalign{\medskip}R_{{2,1}}^{t}&R_{{2,2}}^{t}&R_{{2,3}}^{t}\\\noalign{\medskip}R
_{{3,1}}^{t}&R_{{3,2}}^{t}&R_{{3,3}}^{t}\end {array} \right]
\end{equation}
where for the matrix elements we have explicitly:
\begin{scriptsize}
$$
R_{1,1}^{t}=\left[ \begin {array}{ccc} \sin \left( \pi  \left(
u+2\eta \right)  \right) &0&0\\\noalign{\medskip}0&\sin \left( \pi
u
 \right) {e^{2/3i\pi \eta}}&0\\\noalign{\medskip}0&0&\sin \left(
\pi u \right) {e^{-2/3i\pi \eta}}\end {array} \right]
R_{1,2}^{t}=\left[ \begin {array}{ccc}
0&0&0\\\noalign{\medskip}\sin \left( 2\pi \eta \right) {e^{1/3i\pi
u}}&0&0\\\noalign{\medskip}0&2\,i \sin \left( 2\pi \eta \right)
\sin \left( \pi u \right) {e^{2/3 i\pi  \left( -u-2\eta \right)
}}&0\end {array} \right]
$$
$$
R_{1,3}^{t}=\left[ \begin {array}{ccc}
0&0&0\\\noalign{\medskip}0&0&0\\\noalign{\medskip}\sin \left(
2\,\pi \,\eta \right) {e^{-1/3\,i\pi \,u}}&0&0\end {array} \right]
R_{2,1}^{t}=\left[ \begin {array}{ccc} 0&\sin \left( 2\,\pi \,\eta
\right) {e^{-1/3\,i\pi
\,u}}&0\\\noalign{\medskip}0&0&0\\\noalign{\medskip}-2\,i\sin
 \left( 2\,\pi \,\eta \right) \sin \left( \pi \,u \right) {e^{2/3\,i
\pi \, \left( u+2\,\eta \right) }}&0&0\end {array} \right]
$$
$$
R_{2,2}^{t}=\left[ \begin {array}{ccc} \sin \left( \pi \,u \right)
{e^{-2/3\,i\pi \,\eta}}&0&0\\\noalign{\medskip}0&\sin \left( \pi
\, \left( u+2\, \eta \right)  \right)
&0\\\noalign{\medskip}0&0&\sin \left( \pi \,u
 \right) {e^{2/3\,i\pi \,\eta}}\end {array} \right]
 R_{2,3}^{t}=\left[ \begin {array}{ccc} 0&0&0\\\noalign{\medskip}0&0&0\\\noalign{\medskip}0&\sin \left( 2\,\pi \,\eta \right) {e^{1/3\,i\pi
\,u}}&0\end {array} \right]
$$
$$
R_{3,1}^{t}=\left[ \begin {array}{ccc} 0&0&\sin \left( 2\,\pi
\,\eta \right) {e^{1/3\,i\pi \,u}}\\\noalign{\medskip}2\,i\sin
\left( 2\,\pi \,\eta
 \right) \sin \left( \pi \,u \right) {e^{-2/3\,i\pi \, \left( u+2\,
\eta \right) }}&0&0\\\noalign{\medskip}0&4\,\sin \left( 2\,\pi
\,\eta
 \right) \sin \left( \pi \,u \right) \sin \left( \pi \, \left( u+2\,
\eta \right)  \right) {e^{1/3\,i\pi \, \left( u-2\,\eta \right)
}}&0
\end {array} \right]
$$
$$
R_{3,2}^{t}=\left[ \begin {array}{ccc} 0&-2\,i\sin \left( 2\,\pi
\,\eta \right) \sin \left( \pi \,u \right) {e^{2/3\,i\pi \, \left(
u+2\,\eta \right) }}&0\\\noalign{\medskip}0&0&\sin \left( 2\,\pi
\,\eta \right) {e^{-1/3 \,i\pi \,u}}\\\noalign{\medskip}4\,\sin
\left( 2\,\pi \,\eta \right) \sin \left( \pi \,u \right) \sin
\left( \pi \, \left( u+2\,\eta
 \right)  \right) {e^{-1/3\,i\pi \, \left( u-2\,\eta \right) }}&0&0
\end {array} \right]
$$
$$
R_{3,3}^{t}=\left[ \begin {array}{ccc} \sin \left( \pi \,u \right)
{e^{2/3\,i\pi \,\eta}}&0&0\\\noalign{\medskip}0&\sin \left( \pi
\,u \right) {e^{-2/3 \,i\pi \,\eta}}&0\\\noalign{\medskip}0&0&\sin
\left( \pi \, \left( u+2 \,\eta \right)  \right) \end {array}
\right]
$$
\end{scriptsize}
The explicit expression for the non-standard trigonometric
$sl(N,\mathbb{C})$ R-matrices was calculated in \cite{AHZ}.\\
\indent In the rational case we have the following R-matrix:
$$
\widetilde{R}^{rat}(u)=\left[ \begin {array}{ccc}
R_{{1,1}}^{r}&R_{{1,2}}^{r}&R_{{1,3}}^{r}\\\noalign{\medskip}R_{{2,1}}^{r}&R_{{2,2}}^{r}&R_{{2,3}}^{r}\\\noalign{\medskip}R
_{{3,1}}^{r}&R_{{3,2}}^{r}&R_{{3,3}}^{r}\end {array} \right]
$$
with the following matrix elements
\begin{scriptsize}
$$
R_{1,1}^{r}=\left[ \begin {array}{ccc} 1/2\,{\frac
{u}{\eta}}+1&0&0\\\noalign{\medskip}-2/3\,iu&1/2\,{\frac
{u}{\eta}}&0
\\\noalign{\medskip}-{\frac {16}{27}}\,i{u}^{3}-{\frac {64}{27}}\,i{
\eta}^{2}u-{\frac
{16}{9}}\,i{u}^{2}\eta&-4/3\,{u}^{2}-8/3\,u\eta&1/2 \,{\frac
{u}{\eta}}\end {array} \right] R_{1,2}^{r}=\left[ \begin
{array}{ccc}
0&0&0\\\noalign{\medskip}1&0&0\\\noalign{\medskip}-4/3\,{u}^{2}-8/3\,u\eta&2\,iu&0\end
{array}
 \right]
$$
$$
R_{1,3}^{r}=\left[ \begin {array}{ccc}
0&0&0\\\noalign{\medskip}0&0&0\\\noalign{\medskip}1&0&0\end
{array} \right] R_{2,1}^{r}=\left[ \begin {array}{ccc}
2/3\,iu&1&0\\\noalign{\medskip}-{\frac {8}{9}}\,{u}^{2}-{\frac
{16}{9}}\,u\eta&0&0\\\noalign{\medskip}-{\frac {
32}{81}}\,{u}^{4}-{\frac {32}{27}}\,{u}^{3}\eta-{\frac
{64}{27}}\,{u}^ {2}{\eta}^{2}-{\frac
{256}{81}}\,u{\eta}^{3}&{\frac {8}{27}}\,i{u}^{3} +{\frac
{32}{9}}\,iu{\eta}^{2}+{\frac {16}{9}}\,i{u}^{2}\eta&-2/3\,iu
\end {array} \right]
$$
$$
R_{2,2}^{r}=\left[ \begin {array}{ccc} 1/2\,{\frac
{u}{\eta}}&0&0\\\noalign{\medskip}0&1/2\,{\frac
{u}{\eta}}+1&0\\\noalign{\medskip}{ \frac {8}{9}}\,i{u}^{3}+{\frac
{32}{27}}\,i{\eta}^{2}u+{\frac {16}{9}}
\,i{u}^{2}\eta&0&1/2\,{\frac {u}{\eta}}\end {array} \right]
R_{2,3}^{r}=\left[ \begin {array}{ccc}
0&0&0\\\noalign{\medskip}0&0&0\\\noalign{\medskip}-2/3\,iu&1&0\end
{array} \right]
$$
$$
R_{3,1}^{r}=\left[ \begin {array}{ccc} {\frac
{16}{27}}\,i{u}^{3}+{\frac {64}{27}}\,i{\eta}^{2}u+{\frac
{16}{9}}\,i{u}^{2}\eta&-4/3\,{u}^{2}-8/3\,u\eta
&1\\\noalign{\medskip}-{\frac {32}{81}}\,{u}^{4}-{\frac
{32}{27}}\,{u} ^{3}\eta-{\frac {64}{27}}\,{u}^{2}{\eta}^{2}-{\frac
{256}{81}}\,u{\eta }^{3}&-{\frac {8}{9}}\,i{u}^{3}-{\frac
{16}{9}}\,i{u}^{2}\eta-{\frac {
32}{27}}\,i{\eta}^{2}u&2/3\,iu\\\noalign{\medskip}-{\frac
{1024}{243}} \,{u}^{2}{\eta}^{4}-{\frac
{512}{243}}\,{u}^{3}{\eta}^{3}-{\frac {128}
{243}}\,{u}^{5}\eta-{\frac {128}{729}}\,{u}^{6}-{\frac
{256}{243}}\,{u }^{4}{\eta}^{2}-{\frac
{4096}{729}}\,u{\eta}^{5}&-{\frac {32}{81}}\,i{ u}^{5}+{\frac
{512}{81}}\,i{\eta}^{4}u-{\frac {64}{81}}\,i{u}^{4}\eta+ {\frac
{256}{81}}\,i{u}^{2}{\eta}^{3}&{\frac {8}{27}}\,i{u}^{3}-{ \frac
{32}{27}}\,i{\eta}^{2}u\end {array} \right]
$$
$$
R_{3,2}^{r}=\left[ \begin {array}{ccc}
-4/3\,{u}^{2}-8/3\,u\eta&-2\,iu&0\\\noalign{\medskip}-{\frac
{8}{27}}\,i{u}^{3}-{\frac {32}{9}}\,iu{ \eta}^{2}-{\frac
{16}{9}}\,i{u}^{2}\eta&0&1\\\noalign{\medskip}-{ \frac
{512}{81}}\,i{\eta}^{4}u-{\frac {256}{81}}\,i{u}^{2}{\eta}^{3}+{
\frac {64}{81}}\,i{u}^{4}\eta+{\frac {32}{81}}\,i{u}^{5}&-{\frac
{32}{ 27}}\,{u}^{4}-{\frac {32}{9}}\,{u}^{3}\eta-{\frac
{64}{9}}\,{u}^{2}{ \eta}^{2}-{\frac
{256}{27}}\,u{\eta}^{3}&4/3\,{u}^{2}+8/3\,u\eta
\end {array} \right]
$$
$$
R_{3,3}^{r}=\left[ \begin {array}{ccc} 1/2\,{\frac
{u}{\eta}}&0&0\\\noalign{\medskip}2/3\,iu&1/2\,{\frac {u}{\eta}}&0
\\\noalign{\medskip}-{\frac {8}{27}}\,i{u}^{3}+{\frac {32}{27}}\,i{
\eta}^{2}u&4/3\,{u}^{2}+8/3\,u\eta&1/2\,{\frac {u}{\eta}}+1
\end {array} \right]
$$
\end{scriptsize}
\\
\\
\begin{Large}\textbf{Acknowledgments} \end{Large}\\
\\
The author is grateful to M. Olshanetsky, A. Levin, and A. Zotov
for fruitful discussions and interest to this work. The work was
partly supported by RFBR grant 09-02-00393, RFBR grant
06-01-92054-$KE_{a}$, RFBR-CNRS grant
09-01-93106,  and grant for support of scientific schools NSh-3036.2008.2. The work was
also supported in part by the "Dynasty" Foundation.  \\
\\
\\

\end{document}